\documentclass[twocolumn,nofootinbib]{revtex4-1}

\usepackage{hyperref}  
\usepackage{amsmath}  
\usepackage{amsfonts} 
\usepackage{graphicx} 

\begin{document}


\title{Full non--LTE spectral line formation II. Two--distribution
  radiation transfer with coherent scattering in the atom's frame}

\author{Fr\'ed\'eric Paletou} \email{frederic.paletou@univ-tlse3.fr}
\affiliation{Universit\'e de Toulouse, Observatoire
  Midi--Pyr\'en\'ees, Cnrs, Cnes, Irap, F--31400 Toulouse, France}

\author{Malali Sampoorna} \email{sampoorna@iiap.res.in}
\affiliation{Indian Institute of Astrophysics, Koramangala, Bengaluru
  560034, Karnataka, India}

\author{Christophe Peymirat} \email{christophe.peymirat@univ-tlse3.fr}
\affiliation{Universit\'e de Toulouse, Facult\'e des Sciences et
  d'Ing\'enierie, F--31062 Toulouse cedex 9, France}


\date{\today}

\begin{abstract}
In the present article, we discuss a numerical method of solution for
the so-called ``full non-LTE'' radiation transfer problem, basic
formalism of which was revisited by Paletou \& Peymirat (2021; see
also Oxenius 1986). More specifically, usual numerical iterative
methods for non-LTE radiation transfer are coupled with the
above-mentioned formalism. New numerical additions are explained in
detail. We benchmark the whole process with the standard non-LTE
transfer problem for a two-level atom with Hummer's (1962, 1969)
$R_{\rm I-A}$ partial frequency redistribution function. We finally
display new quantities such as the spatial distribution of the
velocity distribution function of excited atoms, that can only be
accessed to by adopting this more general frame for non-LTE radiation
transfer. \end{abstract}



   \maketitle

%
\section{Introduction}

Oxenius (1986) formulated the so-called ``full non-LTE'' radiation
transfer problem, wherein the distribution of photons as well as the
massive particles in a stellar atmosphere may in general \emph{both}
deviate from their equilibrium distributions. Paletou \& Peymirat
(2021, hereafter PP21) revisited this formalism and rediscussed some
basic elements using standard notations prevalent in this field of
research. In PP21, we however, limited ourselves to the more detailed
statistical equilibrium equations for the simplest case of a
``two-distribution'' model.

The present study is devoted to the \emph{coupling} of that formalism
with usual numerical methods used for radiation transfer. The set and
sequence of the new quantities required for our computations are
outlined in §2 hereafter, in the continuation and sometimes the
generalization of relationships previously discussed in PP21.

The validation of these new elements, dealing directly with the
coupling of the frequency dependence of the radiation and the
velocities of the atoms scattering light, is first made for the case
of coherent scattering \emph{in the atom's frame}. We can indeed show
-- see §3 -- that the latter assumption for our two-distribution
problem makes it equivalent to a two-level atom problem with the
standard angle-averaged partial frequency redistribution (PRD)
function $R_{\rm I-A}$ for isotropic scattering introduced by Hummer
(1962).

Useful details about the very numerical implementation of our new
computations are exposed in §4 and a simple iterative scheme is
described in §5. We fully rely on a nowadays quite usual
short-characteristics formal solver, mostly used for iterative methods
such as accelerated $\Lambda$-iteration (hereafter ALI), whose
material was developed and made available by
us\footnote{\url{https://hal.archives-ouvertes.fr/hal-02546057}} in
Python (see e.g., Lambert et al. 2016 and references therein).

A first, indispensable validation is presented in \S6, by reproducing
results for standard PRD-$R_{\rm I-A}$ of Hummer (1969). It is
however shown that, the two-distribution model provides additional
physical quantities, which \emph{cannot} be accessed to using the
classical non-LTE framework, such as the spatial distribution of the
velocity distribution function (hereafter VDF; see §6 and 8) of the
excited atoms.

Sections 7 and 8 therefore present, respectively, a preliminary
exploration of the effects of velocity-changing elastic collisions,
and new computations for a strongly illuminated finite slab case.

We finally discuss the various perspectives of this study.

\section{Coupling to radiation transfer}

The implementation of our kinetic approach together with radiation
transfer requires, in a first place, two major modifications of
existing numerical radiation transfer tools.

In the following, and after Paletou \& Peymirat (2021), we shall adopt
both the reduced frequency $x$, and the normalized atomic velocity
$u$. The reduced frequency is usually defined in radiation transfer as
$x=(\nu-\nu_0)/\Delta \nu_D$ that is the frequency shift from the line
center normalized to the Doppler width $\Delta \nu_D=(\nu_0/c)
v_{th.}$, where $c$ is the speed of light; while $u$ is the
modulus of the atomic velocity normalized to the ``most probable
velocity'' $v_{th.} = \sqrt{2 k T/M}$, with $k$ being the
Boltzmann constant, $T$ the temperature, and $M$ the mass of the atom.

The first requirement is the computation of the scattering integral
$J_{12}$ defined as:

\begin{equation}
J_{12}(\vec{u},\tau) = \oint{ { {d\Omega} \over {4 \pi} } } \int{
    \delta(x - \vec{u} \cdot \vec{\Omega}) I(x, \vec{\Omega}, \tau) dx} \, ,
\end{equation}
in the case of coherent scattering in the atomic frame, where $\delta$
is the Dirac distribution for the a priori known absorption profile
and $I(x, \vec{\Omega}, \tau)$ the usual specific intensity. The
latter is computed at every frequency $x$, photon propagation
direction $\vec{\Omega}$, and optical depth $\tau$ using a ``classic''
short characteristic based formal solver (see e.g. Lambert et
al. 2016, and associated resources). Once this quantity is available,
our modified formal solver calls a specific function which performs
the angular integration over the Dirac distribution $\delta(x -
\vec{u} \cdot \vec{\Omega})$, adapted from Sampoorna et al. (2011; see
also \S4).

In a second step, one computes $\bar{J}_{12} (\vec{u},\tau) = {J}_{12}
(\vec{u},\tau)/ {\mathcal B}_W$ with ${\mathcal B}_W$ denoting the Planck
function in the Wien limit, and:

\begin{equation}
{\cal{J}}_{12}(\tau) = \int_{\vec{u}}{ \bar{J}_{12}(\vec{u}, \tau) 
f^{M}(\vec{u}) d^3 \vec{u}} \, ,
\end{equation}
with $f^{M}(\vec{u})={\rm e}^{-\vec{u}\cdot\vec{u}}/\pi^{3/2}$ for the
Maxwellian velocity distribution. Here $d^3\vec{u}=u^2du d\Omega_u$
wherein $d\Omega_u=\sin\theta_u d\theta_u d\phi_u$, with $\theta_u$
and $\phi_u$ denoting the polar angles of the normalized atomic
velocity vector $\vec{u}$ about the atmospheric normal.

Once we have computed these two quantities, we can evaluate the
velocity distribution function of the first excited state of the
atom\footnote{In the present study, following PP21 and Oxenius (1986)
we do \emph{not} consider departure of the VDF of the fundamental
state of the atom from Maxwellian. This is a realistic assumption in
the so-called ``weak radiation field regime'' when stimulated emission
can be neglected, leading to a ``natural population'' of the lower
level. This is also the case for an atomic ground level of infinite
lifetime.} using, as defined in PP21:

\begin{widetext}
\begin{equation}
f_2(\vec{u}, \tau) = \left[ \frac{\zeta}{1+\zeta} +
    \left(\frac{1}{1+\zeta}\right) \frac{\varepsilon + (1-\varepsilon)
	\bar{J}_{12} (\vec{u}, \tau)} {\varepsilon + (1-\varepsilon)
	{\cal{J}}_{12}} \right] f^{M}(\vec{u}) \, ,
\end{equation}
\end{widetext}
at every depth in the atmosphere. In this expression, $\varepsilon$ is
the usual collisional destruction probability of \emph{standard}
non--LTE radiation transfer. However, another quantity $\zeta$ appears
now, which characterizes the amount of \emph{velocity-changing elastic
collisions} defined in Eq.\,(27) of PP21 (note again that in PP21 the
depth dependence of the different quantities such as $\bar{J}_{12}$,
$f_2$ etc. were omitted, as radiative transfer was not considered in
detail at this stage).

The next critical quantity is the computation of the emission profile,
which is defined, at a given optical depth $\tau$ in our (1D)
atmosphere as:

\begin{equation}
\psi(x, \tau) = \oint{ { {d\Omega} \over {4 \pi} } } \int_{\vec{u}}{
\delta(x - \vec{u} \cdot \vec{\Omega}) f_2(\vec{u}, \tau) d^3 \vec{u}} \, ,
\end{equation}
where $f_2$ is the VDF of the \emph{excited} atoms scattering light,
computed using Eq.\,(3). At this stage, we apply Eq.\,(A.1) of PP21 to
first perform the integral over $d\Omega$ and then perform the
integral over $d\Omega_u$ to give:

\begin{widetext}
\begin{equation}
   \psi(x, \tau) = \frac{2}{\sqrt{\pi}} \int_{\lvert x
     \rvert}^{\infty}{ \left[ \frac{\zeta}{1+\zeta} +
       \left(\frac{1}{1+\zeta}\right) \frac{\varepsilon +
         (1-\varepsilon) \tilde{J}_{12} (u, \tau)} {\varepsilon +
         (1-\varepsilon) {\cal{J}}_{12}} \right] u e^{-u^2} du } \, ,
\end{equation}
\end{widetext}
where:

\begin{equation}
\tilde {J}_{12} (u, \tau) = \oint \bar {J}_{12}(\vec{u},\tau) d\Omega_u\,.
\end{equation}
Clearly, the emission profile given in Eq.\,(5) above represents the
generalization of the same analytical angular integration result in
Eq.\,(42) of PP21 which however was written only for the $\zeta = 0$
and $\varepsilon = 0$ cases.

Once all the quantities defined by Eqs.\,(1)--(5) have been evaluated,
one may then compute the source function. Hereafter we shall use the
below expression, not given explicitly in PP21 though, for the source
function:

\begin{equation}
  S(x, \tau) = \left[ \varepsilon + (1 - \varepsilon)
    {\cal{J}}_{12}(\tau) \right] \left[ {{\psi(x, \tau)} \over
      {\varphi(x)}} \right] \, ,
\end{equation}
where $\varphi$ is the so-called Doppler absorption profile. It
is quite straightforward to establish, and this is indeed the very
expression of the source function that we used in the present study
(note that this expression remains \emph{unchanged} when $\zeta \neq
0$).

\section{Equivalence with Hummer's $R_{\rm I-A}$ redistribution}

It is easy to show that a ``two-distribution'' model assuming a
Maxwellian distribution for $f_1$, characterizing the fundamental
level, and coherent scattering in the atom's frame is equivalent to a
standard PRD model for a two--level atom problem considering Hummer's
angle-averaged (and isotropic scattering) $R_{\rm I-A}$
redistribution. It therefore provides a \emph{critical} benchmark for
our modified numerical tools.

A way to verify this equivalence is by using the developed expression
of the emission profile given by the combination of Eqs.\,(3) and (4)
together with Eq.\,(7). This should also be considered for $\zeta=0$
since this new parameter is \emph{not} relevant to standard PRD using
Hummer's redistribution functions.  Then, one may split in two the
expression of the emission profile with a first part which goes like
$\varepsilon f^M(\vec{u})$, and another one which implies
$(1-\varepsilon) \bar{J}_{12} (\vec{u},\tau) f^M(\vec{u})$ -- besides
the common denominator which is independent from $\vec{u}$. The first
part, after integrations, will lead to the Doppler absorption profile
$\varphi$, since it is the convolution of the Dirac function
characterizing coherent scattering in the atom's frame with the
Maxwellian VDF $f^M(\vec{u})$. But most interesting is however the
second part involving $\bar{J}_{12}$. For this second part we go back
to real frequencies $\xi$ (in the atomic frame) and $\nu$ (in the
observer's frame), from $x$ and the Doppler transform (see also
notations adopted in PP21). Now the Dirac function appearing in
Eq.\,(4) becomes $\delta(\xi - \nu_0)$, while that in Eq.\,(1) when
substituted in Eq.\,(4) becomes $\delta(\xi^\prime - \nu_0)$. Thus,
the second part of the emission profile contains the product of two
Dirac distributions, namely $\delta(\xi - \nu_0)\delta(\xi' -
\nu_0)$. One can also rewrite this combination as: $\delta(\xi -
\nu_0)\delta(\xi' - \xi)$ which is indeed the \emph{atomic}
redistribution function $r_{\rm I}$ (see e.g., Hubeny \& Mihalas,
2014). This has also been discussed in \S4 of Borsenberger et
al. (1987).

After integration over velocities along the Maxwellian VDF, and after
angular integration, we finally recover the more usual form of the
standard PRD, frequency-dependent source function (normalized to the
Planckian):

\begin{widetext}
\begin{equation}
  S(x, \tau) = \varepsilon + (1 - \varepsilon) \oint{ { {d\Omega'}
      \over {4 \pi} } } \int_{x'} { \left[ \frac{R_{\rm
          I-A}(x',x)}{\varphi(x)} \right] I(x', \Omega', \tau) dx'} \, .
\end{equation}
\end{widetext}
And the solution of such a problem can easily be computed using the
methods proposed by Paletou \& Auer (1995).

\section{Numerical implementation}

Our numerical implementation of the new, full non-LTE problem relies
on modifications brought to a nowadays classic short-characteristics
(hereafter SC) based formal solver (see e.g. Lambert et al. 2016, and
references therein). This, together with various iterative schemes
which can be set on that basis constitutes a more efficient way, both
fast and accurate, as well as easy to develop on, to address the
problem than what was made previously by Borsenberger et al. (1986,
1987) and Atanackovi$\check{\rm c}$ et al. (1987).

The main numerical problem now, as compared to the standard non-LTE
problem, is to properly evaluate expressions given by Eqs. (1) and
(4). First is the angular integration for making $J_{12}$. It consists
in using a quadrature in polar angle and azimuth $(\theta, \phi)$
\emph{both} for the ray direction and for the atomic velocity. Then,
one may write: $ \vec{u} \cdot \vec{\Omega} = \gamma u$ where:

\begin{equation}
  \gamma = \cos(\theta_r)\cos(\theta_u) +
  \sin(\theta_r)\sin(\theta_u)\cos(\phi_r - \phi_u) \, .
\end{equation}
Here indices $r$ and $u$ are respectively associated with the ray
and the atomic velocity directions.

Once $\gamma$ has been computed for every couple of polar angle and
azimuth, the specific intensity is interpolated in $x$ in order to
estimate the $I(x=\gamma u)$ quantity that will contribute to the
integral leading to $J_{12}(\vec{u},\tau)$, after a first integration
in frequency along the atomic absorption profile. A mere linear
interpolation was used together with our identical $x$ and $u$ grids,
hereafter spanning 6 Doppler width in $x$, with $\Delta u = \Delta x =
0.1$. To achieve this very task, we use a dedicated function which is
called in from the usual SC formal solver, once the specific intensity
is available at all frequencies, at a given depth and direction.

Practically, we used a 10-point regularly spaced quadrature for $\phi$
in the $[0, 2\pi]$ domain, together with Gauss--Legendre nodes for the
direction cosines usually defined in radiation transfer as: $\mu =
\cos(\theta_r)$. Most computations used 6 nodes for $\mu$, leading
therefore to 60 distinct couples $(\theta, \phi)$. Note also that our
new formal solver was designed for considering full frequency
\emph{and} angular dependence of the source function.

The numerical calculation of ${\cal{J}}_{12}$ is straightforward. Here
for the integration over $d\Omega_u$ we use the same angular
quadrature as mentioned above. The integration over the modulus of the
atomic velocity $u$ can be done either using a simple trapezoidal rule
or a Gauss--Hermite quadrature.

Then follows another similar numerical integration over atomic
velocities, leading to the \emph{self-consistent} emission profile,
according to Eq.\,(5). It is repeatedly done further using a basic
trapezoidal rule. It was obviously being tested setting $f_2 \equiv
f^{M}$, for which case we easily recover the usual thermal, or Doppler
profile:

\begin{equation}
  \varphi(x) = \frac{1}{\sqrt{\pi}}  e^{-x^2}  \, ,
\end{equation}
accurately i.e., with a relative error better than 1\% over the
spectral domain we used for the radiation transfer problem.

Moreover, every additional numerical calculations previously listed
have been tested for the recovery of the well-known solution of the
two-level atom with complete redistribution in frequency (hereafter
CRD) problem.

\section{A simple iterative scheme}

A first validation was indeed to run the whole process with the
modified formal solver now also computing $\bar{J}_{12}
(\vec{u},\tau)$ setting $f_2 \equiv f^M$, so that we could recover the
usual CRD solution, for a Doppler absorption profile. Our angular
quadrature with 10 azimuths and 6 Gauss--Legendre nodes for the
$\mu$'s guarantees a CRD--like solution within 1.5\% \emph{maximum}
relative error throughout a semi-infinite atmosphere.

Second, for benchmarking with the standard PRD case using $R_{\rm
  I-A}$ redistribution, we used a very simple iterative scheme
consisting, once the CRD solution has been computed using standard
ALI, in a mere computation and successive updates of $J_{12}$ and
${\cal{J}}_{12}$, then $f_2$ and $\psi$ and update the source function
$S$ before moving to the next iteration. The latter process is somewhat
comparable to $\Lambda$-iteration in the sense that it consists in the
simplest possible iterative process one could implement. A similar
process was for instance successfully used by Paletou et al. (1999)
for polarized radiative transfer in 2D geometry. The same numerical
strategy allows us to consider also cases for which $\zeta \ne 0$ (see
§7 hereafter) \emph{without} any difficulty.

\begin{figure}[]
  \includegraphics[width=10 cm, angle=0]{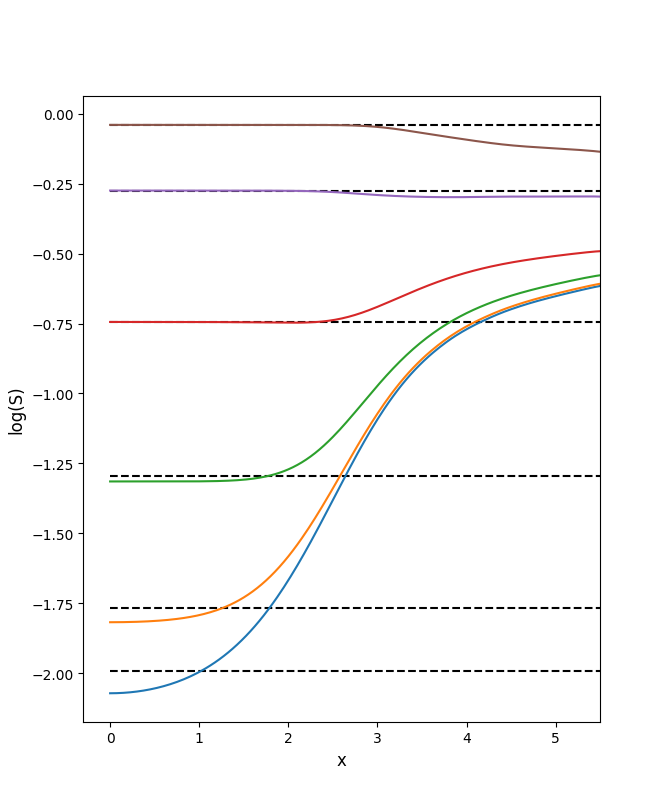}
  \caption{Variation with frequency of the normalized source function,
    for different values of the optical depth at line center across
    the atmosphere. Dashed lines indicate the (constant with
    frequency) CRD values at $\tau = 0, \,1, \,10, \,100,\,10^3,
    \,10^4$ for comparison, where both $S_{\rm CRD}$ and $S(x=0)$
    increase with $\tau$. It satisfactorily reproduces the standard
    PRD results using Hummer's $R_{\rm I-A}$ (see also Hummer 1969,
    Fig.\,1c).}
  \label{Fig1}
\end{figure}

\section{Benchmarking against Hummer's $R_{\rm I-A}$}

\begin{figure}[]
  \includegraphics[width=9.5 cm, angle=0]{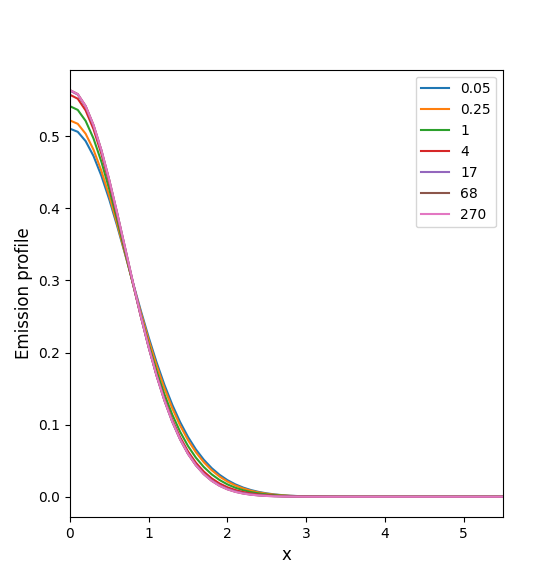}
  \caption{Dependence of the emission profiles $\psi(x, \tau)$ on
    various optical depths across the atmosphere (see the top-right
    frame giving the correspondence between $\tau$ at line center and
    the color of the relevant profile). This distribution is computed
    self-consistently with the radiation, without the need of any a
    priori given redistribution function.}
  \label{Fig2}
\end{figure}

Our first task then has been to reproduce the $S(x, \tau)$ results of
the original Hummer (1969; his Fig. 1c) publication. It is an obvious
comparison to be made, which was not conducted in earlier
studies. They were obtained for a 1D, semi-infinite, plane parallel
atmosphere of total optical thickness at line center $\tau=10^6$,
$\varepsilon=10^{-4}$ and $\zeta=0$. In our computations, we used 5
points per decade in order to cover this range. Our solutions for
$S(x,\tau)$ are displayed in Fig.\,(1) where increasing values around
line core ($|x| < 2$) correspond to successive optical depths of the
order of: $\tau = 0, \,1, \,10, \,100,\,10^3, \,10^4$. Successive
dashed lines mark the (frequency independent) CRD values at the same
optical depths, for comparison.

\begin{figure}[]
  \includegraphics[width=9.5 cm, angle=0]{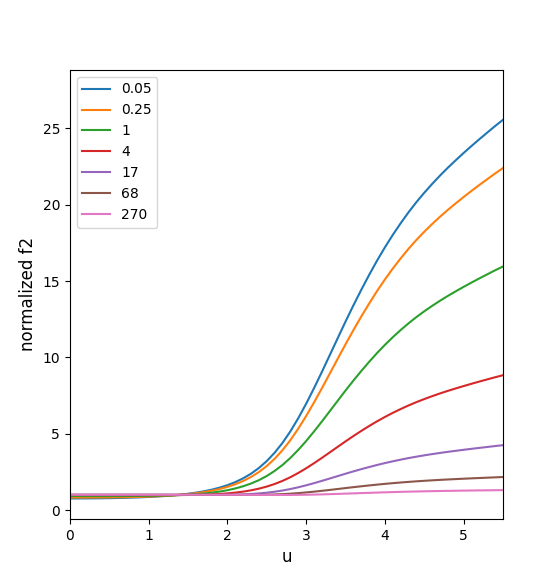}
  \caption{Deviations from Maxwellian illustrated by ratios $f_2(u,
    \tau)/f^M (u)$ at various optical depths across the atmosphere
    using the same convention as in Fig.\,(2). It shows the
    ``overpopulation'' of $f_2$ at large $u$'s. This very information
    cannot be accessed to, using standard non--LTE approaches.}
  \label{Fig3}
\end{figure}

Even using our simple iterative scheme, we recover easily the Hummer
(1969) ``historic'' solutions. We could also check that the relative
error on $S(x, \tau)$ between our new scheme vs. similar results
obtained with a ``frequency by frequency'' (hereafter FBF; see Paletou
\& Auer, 1995) for $R_{\rm I-A}$ never exceeds 4\% with a more typical
mean value around 1.3\% (for this validation, we used FBF with a
6-node Gauss-Legendre angular quadrature).

To achieve this level of accuracy, we compute the CRD solution using
the ALI-iteration until the maximum relative error on the source
function was better than $10^{-4}$. The new cycle was then iterated up
to a maximum relative error on the frequency dependent source function
of $3 \times 10^{-4}$.

In Fig.\,(2), we display the emission profile $\psi(x, \tau)$ for
different optical depths $\tau$ across the atmosphere. These are
``naturally'' computed using our approach, and \emph{without} the need
of any a priori given redistribution function. The \emph{only}
assumptions we rely on are: (1) that the VDF $f_1$ of the atoms in
their ground state is Maxwellian, and (2) that the \emph{atomic}
absorption profile is a priori given, in the present case of coherent
scattering by a Dirac function centered at the frequency $\nu_0$ of
the model-spectral line. Then all relevant quantities, down to the
velocity distribution function of the excited atoms at every depth
into the atmosphere, are computed \emph{consistently with} the
successive evaluation of the radiation field.

Indeed, we can display, as shown in Fig.\,(3) a more original sample
of the ratios $f_2(u, \tau)/f^M(u)$ for different optical depths, as
considered in Fig.\,(2). In the present paper, we consider the
emission profile $\psi(x,\tau)$ and thereby the source function
$S(x,\tau)$ to be independent of the polar angles of the radiation
field. Therefore, we prefer to illustrate the normalized VDF of the
excited atom that depends only on the modulus of velocity $u$. This is
obtained by integrating $f(\vec{u},\tau)$ given by Eq.\,(3) over
atomic velocity directions, namely $d\Omega_u$. Important deviations
from Maxwellian can be identified, typically for $u > 2$, and close to
the non-illuminated surface of the semi-infinite atmosphere. We
recover easily, using our new numerical procedures, the
``overpopulation'' of $f_2$ at large $u$'s already put in evidence by
Borsenberger et al. (1987).

Such a material is \emph{only} accessible using the Oxenius-like
formalism that we adopted. Note also that the \emph{de facto}
neglected potential effects of velocity-changing elastic collisions
can only be addressed, and studied in that theoretical frame (see
details in PP21 and \S7 hereafter).  More generally, such additional
information could be very valuable for the detailed coupling between
the radiation transfer problem and any other physical processes that
would take place within an atmosphere, and for which the very
knowledge of the various VDF's of contributing elements, at different
excitation and ionization stages would be critical.

\section{Non-zero velocity-changing elastic collisions}

To the best of our knowledge, such preliminary computations were only
addressed by Atanackovi$\check{\rm c}$ et al. (1987) so far.

We now go a bit further by illustrating in Fig.\,(4) the dependence of
the source function at $\tau \approx 1$ on the velocity-changing
elastic collisions parameterized as $\zeta$. The model-atmosphere used
for this computation is identical to the one adopted in \S6, but we
have made vary $\zeta$ from $0$ to $50$. As expected, the source
function $S(x,\tau)$ ranges between the standard PRD-$R_{\rm I-A}$
solutions when $\zeta=0$, and the frequency independent CRD values
(shown as black dashed line) for increasing values of $\zeta$.

\begin{figure}[]
  \includegraphics[width=9.9 cm, angle=0]{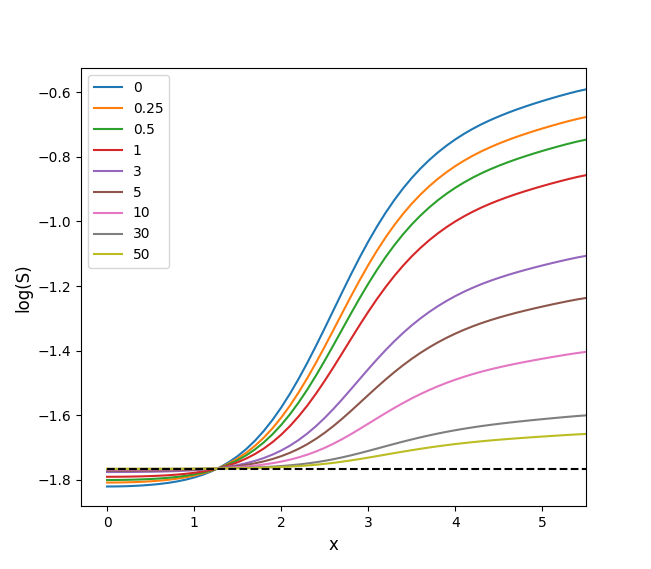}
\caption{Variation with frequency of the normalized source function at
  $\tau \approx 1$, for different values of the velocity-changing
  elastic collision parameter $\zeta$ (indicated in the top-left
  frame). The model-atmosphere is the same as the one considered in
  \S6 and Hummer (1969). As expected, successive solutions range
  between standard PRD-$R_{\rm I-A}$ values (blue) and the limit of
  the CRD solution (dashed line) for increasing values of $\zeta$.}
  \label{Fig4}
\end{figure}

As also expected, the numerical problem becomes \emph{easier} for
increasing values of $\zeta$. Effects of velocity-changing elastic
collisions will be discussed in more details in another devoted study.

\section{A finite slab case}

Among numerous applications, we are particularly interested in the
radiative modelling of isolated and illuminated finite slabs. As an
example of expected effects, we simulated a 1D plane parallel
horizontal slab strongly irradiated asymmetrically, only from
below. This mimics the radiative modelling of a ``cold filament''
suspended above a stellar disk (see e.g., Paletou 1997).

We used a 33 points optical depth grid, logarithmically spaced away
from \emph{both} open surfaces (using an initial $\delta \tau=0.01$),
and symmetric around a midslab depth set at $\tau_{1/2}=500$. For this
example we also set $\varepsilon=10^{-4}$ and $\zeta=0$. External
illumination is applied \emph{only} at the bottom surface, using a
flat profile of normalized intensity $I_{\rm ext.}=3$.

Figure (5) displays values of the normalized VDF $f_2(u,\tau)/f^M(u)$
for optical depth values, counted from the top surface, of $\tau=0.58$
and $\tau=1.15$ i.e., around the critical value of 1, and at
midslab. The quite strong external illumination that we applied
generates very significant departures of $f_2$ from Maxwellian,
especially at $u > 2$ again, but larger than what we already
identified for the semi-infinite atmosphere case.

\begin{figure}[]
  \includegraphics[width=9.5 cm, angle=0]{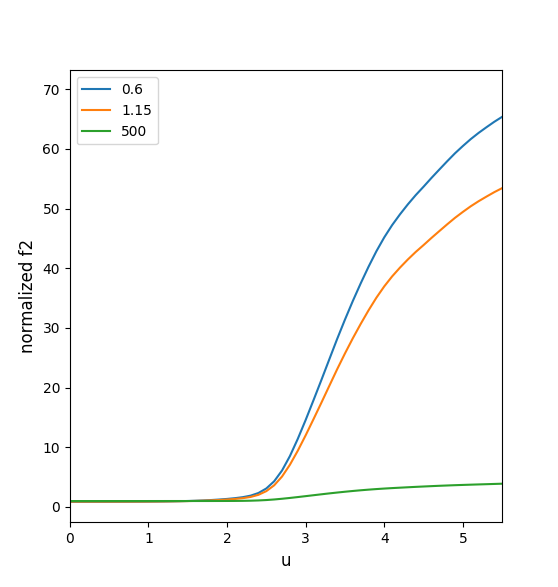}
  \caption{Deviations from Maxwellian illustrated by the changes of
    the normalized $f_2$ at various optical depths (mentioned in the
    top-left frame) across an asymmetrically and strongly illuminated
    finite slab of total depth $\tau=10^3$. Largest amplitudes are for
    these two values around $\tau=1$, while smaller but significant
    deviations are still noticeable at midslab ($\tau=500$).}
  \label{Fig5}
\end{figure}

\section{Discussion and conclusion}

A first critical evaluation of new numerical procedures in radiative
transfer for the solution of the ``full non--LTE'' problem have been
conducted. They were validated by reproducing both CRD and standard
PRD with Hummer's $R_{\rm I-A}$ (Hummer 1962, 1969). This will allow
us to now move forward in several directions.

After the very first computations shown here, we shall also be able to
evaluate further, in more details, the additional effects of potential
velocity-changing \emph{elastic} collisions with different set of
``classic'' parameters. Also \emph{finite} slab models, with different
conditions of external illumination will be considered. Such cases
lead to more significant departures from Maxwellian than what happens
for semi-infinite slabs, as indicated by our preliminary filament-like
computation. Relevant astrophysical ``objects'' should range from
solar prominences to circumstellar environements for instance.

The next obvious step is to modify the atomic absorption description
for the more realistic case of natural broadening of the upper level
of the transition. It would go beyond the previous studies of
Borsenberger et al. (1987) and Atanackovi$\check{\rm c}$ et al. (1987)
which were limited to ``pure'' Doppler broadening. Therefore, a
Lorentzian profile for atomic absorption will be used. This will also
require to implement modifications to the original formalism of
Oxenius for that case, following rather Bommier's (1997) approach, as
discussed in PP21.  This will be suitable for a preliminary study of
resonance lines such as Lyman $\alpha$ of H\,{\sc i} for instance. The
successive computations of Voigt-like profiles at every iterative
step, according to the departures of $f_2$ to Maxwellian, and
throughout the whole atmosphere will certainly benefit from the
numerical scheme proposed by Paletou et al. (2020). This may also
require us to implement and validate a more robust iterative scheme,
more likely inspired by the so-called FBF scheme proposed by Paletou
\& Auer (1995).

Then we shall proceed with the consideration of at least an additional
distribution, either for another excited state or for free
electrons. This should lead to an alternative, less heuristic,
approach to the so-called ``cross-redistribution'' model for the
multi-level atoms case (see e.g., Milkey et al. 1975, Hubeny \& Lites
1995, Sampoorna \& Nagendra 2017).
   
\begin{acknowledgements}
  M.S. acknowledges the support from the Science and Engineering
  Research Board (SERB), Department of Science and Technology,
  Government of India via a SERB-Women Excellence Award research grant
  WEA/2020/000012.
\end{acknowledgements}

\end{document}